\documentclass[aps,prb,twocolumn,superscriptaddress,showpacs,amsmath,amssymb,footinbib,longbibliography]{revtex4-2}
\usepackage[english]{babel}
\usepackage{amssymb,amsbsy,amsmath}
\usepackage{graphicx}
\usepackage{dcolumn}
\usepackage{bm}
\usepackage{subfigure}
\usepackage{color}
\usepackage{textcomp}
\usepackage[colorlinks,bookmarks=false,citecolor=blue,linkcolor=red,urlcolor=blue]{hyperref}
\usepackage[english]{babel}

\def\ket#1{\, | \, {#1} \, \rangle}
\newcommand{\braket}[2]{\langle \, {#1} \, | \, {#2} \, \rangle}

\begin{document}
\title{Magnetism of the $s=1/2$ $J_1$-$J_2$ square-kagome lattice antiferromagnet}

\author{Johannes Richter}
\email{Johannes.Richter@physik.uni-magdeburg.de}
\affiliation{Institut f\"ur Physik, Universit\"at Magdeburg, P.O. Box 4120, D-39016 Magdeburg, Germany}
\affiliation{Max-Planck-Institut f\"{u}r Physik Komplexer Systeme,
        N\"{o}thnitzer Stra{\ss}e 38, D-01187 Dresden, Germany}
\author{J\"urgen Schnack}
\email{jschnack@uni-bielefeld.de}
\affiliation{Fakult\"at f\"ur Physik, Universit\"at Bielefeld, Postfach 100131, D-33501 Bielefeld, Germany}

\date{\today}

\begin{abstract}
Over the last decade, the spin-$1/2$  Heisenberg antiferromagnet
 on the square-kagome (SK) lattice  has attracted growing attention
as a model system of highly frustrated quantum magnetism.
A further motivation for theoretical studies of this model comes from the recent discovery of
 SK spin-liquid compounds 
KCu$_6$AlBiO$_4$(SO$_4$)$_5$Cl
[{\it M. Fujihala et al., Nat. Commun. \textbf{11}, 3429 (2020)}] and
Na$_6$Cu$_7$BiO$_4$(PO$_4$)$_4$[Cl,(OH)]$_3$ [{\it O. V. Yakubovich et al.
Inorg. Chem. {\bf 60}, 11450 (2021)}].
The SK  antiferromagnet exhibits two non-equivalent nearest-neighbor bonds $J_1$ and
$J_2$. One may expect that in SK compounds $J_1$ and
$J_2$ are of different strength.   
Here, we present a numerical study of finite systems of  $N=30$, $36$ and $N=42$ 
sites by means of the
finite-temperature Lanczos method.
We discuss the temperature dependence of  the Wilson ratio $P(T)$, the specific heat
$C(T)$,
the entropy $S(T)$, and of the susceptibility $X(T)$ of the $J_1$-$J_2$
SK  Heisenberg antiferromagnet
varying $J_2/J_1$ in a range $0 \le J_2/J_1 \le 4$.
We also discuss the zero-field ground state of the model.
We find indications for a magnetically disordered singlet ground state for 
$0 \le J_2/J_1 \lesssim 1.65$. Beyond  $J_2/J_1 \sim 1.65$ the  singlet ground state 
gives way for a ferrimagnetic ground state
which becomes a stable Lieb ferrimagnet with magnetization $M=N/6$ (UUD state) for $J_2/J_1
\gtrsim 1.83$.
In the region  $0.77 \lesssim J_2/J_1  \lesssim 1.65$ the low-temperature
thermodynamics is dominated by  
a finite singlet-triplet gap filled with low-lying singlet excitations leading
to an exponentially activated low-temperature behavior
of $X(T)$. On the other hand, the low-lying singlets yield 
an extra maximum or a shoulder-like profile below the main maximum in the  $C(T)$
curve.
For   $J_2/J_1 \lesssim 0.7$ the low-temperature thermodynamics is characterized by a
large fraction
of $N/3$  weakly coupled spins  leading to a sizable amount of entropy at very
low temperatures. 
In an applied magnetic field the magnetization process features
plateaus and jumps in a wide range of
$J_2/J_1$.
\end{abstract}


\maketitle

\section{Introduction}
\label{sec-1}
Highly frustrated quantum antiferromagnets on two-dimensional lattices have attracted
an enormous attention over more than three decades, see, e.g., 
\cite{Moe:CJP01,LNP645,LMM:10,Balents_Nature2010,Balents_review2017}.
"Now in the early 2020s, quantum magnetism is a mature field showing no signs of
senescence. To the contrary, there is a tremendous amount of activity studying exotic magnetic phenomena especially 
with strong quantum fluctuations." \cite{Review_top-magnons2022}
Over many years the kagome antiferromagnet (KHAF) 
has been the holy grail in this field.
Quite recently the square-kagome antiferromagnet, the 'little brother' of the kagome
antiferromagnet, has received more appreciation because several magnetic
compounds with square-kagome lattice structure have been found which do not
exhibit magnetic order down to very low temperatures
\cite{FMM:NC20,YSK:IC21,Vasiliev2022,Vasiliev2023}.
The square-kagome lattice
(sometimes also called shuriken or squagome lattice)
\cite{Sidd2001,richter2004squago,richter2009squago,Sakai2013}
is a two-dimensional tiling built of squares and corner-sharing triangles.  The classical ground state
of the square-kagome  Heisenberg antiferromagnet (SKHAF) 
is highly degenerated (classical spin liquid). 
There are two non-equivalent sites A and B as well as two non-equivalent  nearest-neighbor bonds $J_1$ and
$J_2$, see the left inset in
Fig.~\ref{fig_m+}.
The  theoretical study of the quantum model
started 20 years ago
\cite{Sidd2001,SSR:EPJB01,tomczak2003specific,richter2004squago,richter2004-spin-peierls,richter2009squago}.
Already at that time evidence
for the absence of ground-state magnetic order was found
\cite{richter2004squago,richter2009squago}.   

Starting in 2013 the interest in the spin-$1/2$ SKHAF has been growing
as a model system exhibiting a non-magnetic quantum ground
state, magnetization plateaus, flat-band physics near the saturation field and quantum scars 
\cite{Sakai2013,Rousochatzakis2013,derzhko2014square,Rousochatzakis2015,richter2004-spin-peierls,Sakai2015,DRM:IJMP15,Hasegawa2018,
Morita2018,Lugan2019,McClarty2020,PhysRevB.102.241115,Iqbal2021,schmoll2022tensor}.
All these papers were focused on zero-temperature properties.  
Only, in the early paper \cite{tomczak2003specific} specific-heat
data calculated by a simple renormalization group approach were reported.
The thermodynamics of the balanced spin-$1/2$ SKHAF, i.e., $J_1=J_2=J$,  has been studied
quite recently in Ref.~\cite{squago_TD_2022}  using the finite-temperature Lanczos method
(FTLM).
At zero magnetic field  we find that the KHAF and SKHAF
exhibit  a striking
similarity of the temperature profile of $C(T)$, $X(T)$ and  $S(T)$
down to very low temperature $T$.
Thus, for $X(T)$ and  $S(T)$   an almost perfect coincidence
for both models was observed.
For the specific heat there is a perfect agreement of the $C(T)$ data down to $T/J = 0.3$.
 A characteristic feature common in both models is the existence of  low-energy singlet
excitations filling the magnetic spin gap
\cite{richter2009squago,Lech:1997,Waldtmann1998,LSM:PRB19}.
These low-energy singlets yield a low-temperature shoulder below
the major maximum in the $C(T)$ profile \cite{kago42,squago_TD_2022}. We
mention that such a
shoulder
has been observed in a recent experiment on  
the kagome quantum antiferromagnet
YCu$_3$(OH)$_6$Br$_2$[Br$_x$(OH)$_{1-x}$] \cite{PhysRevB.105.L121109}.
The subtle details of the singlet excitations depending on the shape and
the size $N$ of the finite lattices
lead to deviations between the behavior 
of $C(T)$ for both models at very low 
$T$.

Bearing in mind the recent experimental studies on
square-kagome quantum antiferromagnets
\cite{FMM:NC20,YSK:IC21,Vasiliev2022,Vasiliev2023}
and the non-equivalence
of the  nearest-neighbor bonds $J_1$ and $J_2$ we may expect that for the
modeling
of square-kagome compounds
it is natural to consider a spin model with $J_1 \ne J_2$.
Moreover,
the $J_1$-$J_2$ model is interesting in its own right as highly frustrated model that allows to tune
the competition between the bonds.
        
So far only a few papers exist which study the zero-temperature properties
of the $J_1$-$J_2$ model \cite{Rousochatzakis2013,Rousochatzakis2015,Hasegawa2018,Morita2018,Lugan2019} where in
Ref.~\cite{Hasegawa2018} the focus is on the magnetization process of the $J_1$-$J_2$ model with 
only slight deviations from the balanced model, i.e.,  the difference between $J_1$ and $J_2$
is small.
In our paper we will fill the gap of missing nonzero-temperature studies and present FTLM data for the
magnetization $M$, the Wilson ratio $P$, the specific heat $C$, the entropy $S$ and the
uniform magnetic susceptibility $X$ of the $J_1$-$J_2$ SKHAF.
In addition, we will analyze the ground state of the finite lattices used
for the FTLM studies which allows to get a relation between ground-state and
finite-temperature properties of the investigated systems.

The corresponding Heisenberg Hamiltonian  augmented with a Zeeman term is given by 
\begin{equation}
\label{Ham}
H
=
J_1 \hspace*{-2mm} \sum_{<i,j>_1}
{\bf s}_i \cdot {\bf s}_j
+
J_2 \hspace*{-2mm}
 \sum_{<i,j>_2}
{\bf s}_i \cdot {\bf s}_j
+
g \mu_B\, B
\sum_{i}
s^z_i
\ ,
\end{equation}
where ${\bf s}_i^2=s(s+1)=3/4$.
The $J_1$ bonds represent the nearest-neighbor exchange connecting A sites on
the squares, whereas the  $J_2$ bonds represent the nearest-neighbor exchange connecting A
with B sites on
the triangles, see the left inset in Fig.~\ref{fig_m+}.
In what follows we set $J_1=1$.

The paper is organized as follow. In Section~\ref{sec-2} we
introduce our numerical scheme. In Section~\ref{sec-3}
we present our results where in Section~\ref{sec-3a} 
we
briefly discuss the ground-state properties as well as the excitation gaps of the model which may
be relevant
for the interpretation of the low-temperature thermodynamics.
The results for the temperature dependence of the Wilson ratio $P(T)$, the specific heat
$C(T)$,
the entropy $S(T)$ as well as the susceptibility $X(T)$
at zero magnetic field are presented and discussed in
Section~\ref{sec-3b}. Finally, in Section~\ref{sec-3c}     
we discuss the magnetization process in an applied magnetic field.
In the last Section~\ref{sec-4}
we summarize our findings.
In two appendices we show the finite lattices considered in our paper
(App.~\ref{sec-a1}) and provide some additional figures to illustrate
finite-size effects (App.~\ref{sec-a2}).

\section{Calculation scheme}
\label{sec-2}

The magnetic system under consideration is modeled by the spin-$1/2$
Heisenberg Hamiltonian given in Eq.~(\ref{Ham}).
We use the
conservation of the $z$-component of the total spin  
$S^z=\sum_{i} s^z_i$ as well as lattice symmetries, i.e., the Hilbert space splits into
subspaces characterized by the  
eigenvalues of $S^z$ (magnetic quantum number $M$) and of the symmetry operator, see, e.g.,
Refs.~\cite{Lauchli_ED_2011,richter2004starlattice}.
To calculate the ground state we perform Lanczos exact diagonalization in the sector 
$M=0$.
For that we use J\"org Schulenburg's publicly available  package {\it spinpack}
\cite{spin:256,richter2010spin}.

For the FTLM 
scheme we also exploit the package {\it spinpack} as well as the conservation of
$S_z$ and the symmetries to decompose  
the Hilbert space into much smaller subspaces.
The FTLM is meanwhile a well established and accurate approach to
calculate thermodynamic quantities of frustrated quantum spin systems
\cite{kago42,squago_TD_2022,JaP:PRB94,HaD:PRE00,ADE:PRB03,ScW:EPJB10,SuS:PRL12,PrB:SSSSS13,
SuS:PRL13,ScT:PR17,PRE:COR17,PrK:PRB18,OAD:PRE18,IMN:IEEE19,MoT:PRR20,Accuracy2020,SGS:ZNA21,PRL_mag_cryst}.
We do not present a detailed description of the method, rather we will provide the
basics of  the FTLM for convenience.  
Within the FTLM 
the sum over an orthonormal basis in the  partition function is replaced by a much smaller sum over
$R$ random vectors: 
\begin{eqnarray}
\label{Z}
Z(T,B)
&\approx&
\sum_{\gamma=1}^\Gamma\;
\frac{\text{dim}({\mathcal H}(\gamma))}{R}
\sum_{\nu=1}^R\;
\sum_{n=1}^{N_L}\;
e^{-\beta \epsilon_n^{(\nu)}} |\braket{n(\nu)}{\nu}|^2
\ ,
\nonumber \\[-3mm]
\end{eqnarray}
where the $\ket{\nu}$ label  random vectors 
for each symmetry-related orthogonal subspace
${\mathcal H}(\gamma)$ of the Hilbert space with  $\gamma$
labeling   the respective 
symmetry. 
In Eq.~(\ref{Z}), the 
exponential of the Hamiltonian has been replaced by its spectral
representation in a Krylov space
spanned by the
$N_L$ Lanczos vectors starting from the respective random vector
$\ket{\nu}$, where
$\ket{n(\nu)}$ is the $n$-th eigenvector of $H$ in
this Krylov space.

For more information we refer the interested reader to the reviews
\cite{PrB:SSSSS13,PRE:COR17} and to our recent FTLM papers of the KHAF
\cite{kago42} and SKHAF  \cite{squago_TD_2022}.  
A detailed discussion of 
the accuracy of the FTLM can be found 
in Refs.~\cite{kago42} and \cite{Accuracy2020}. 
Based on this knowledge, we chose the
number of random vectors $R$ along the lines of our previous
study \cite{kago42}.

\section{The SKHAF at zero magnetic field}
\label{sec-3}
\subsection{Analysis of the ground state of SKHAF on finite lattices of
$N=30$ and $N=36$ sites}
\label{sec-3a}
The absence of magnetic long-range order for the balanced $s=1/2$ SKHAF ($J_1=J_2$)
was
established by previous studies
\cite{richter2009squago,Rousochatzakis2013,Rousochatzakis2015,Lugan2019,Iqbal2021,schmoll2022tensor}.
The nature of the ground state is still under debate, candidates are a pinwheel valence-bond-crystal 
ground state
\cite{Rousochatzakis2013,Iqbal2021}, a loop-six valence-bond state
\cite{Rousochatzakis2015,schmoll2022tensor}
or a topological nematic spin liquid \cite{Lugan2019}.
The ground-state  phase diagram of the $J_1$-$J_2$ model was studied in 
Ref.~\cite{Lugan2019}  
using a  Schwinger-boson mean field theory 
as well as in Refs.~\cite{Rousochatzakis2013,Rousochatzakis2015}
using a  resonating valence-bond
approach.

Here we present Lanczos exact diagonalization data  for  $N=30$  and $N=36$.	
Note that a brief discussion of the ground state for $N=24$ and $N=30$ was
already given in Refs.~\cite{Rousochatzakis2015,Morita2018}.    
Our ground-state  data will be useful to compare with the Schwinger-boson data  \cite{Lugan2019}
as well as  
for the interpretation of the low-temperature thermodynamics.

\begin{figure}[ht!]
\centering
\includegraphics*[clip,width=1.1\columnwidth]{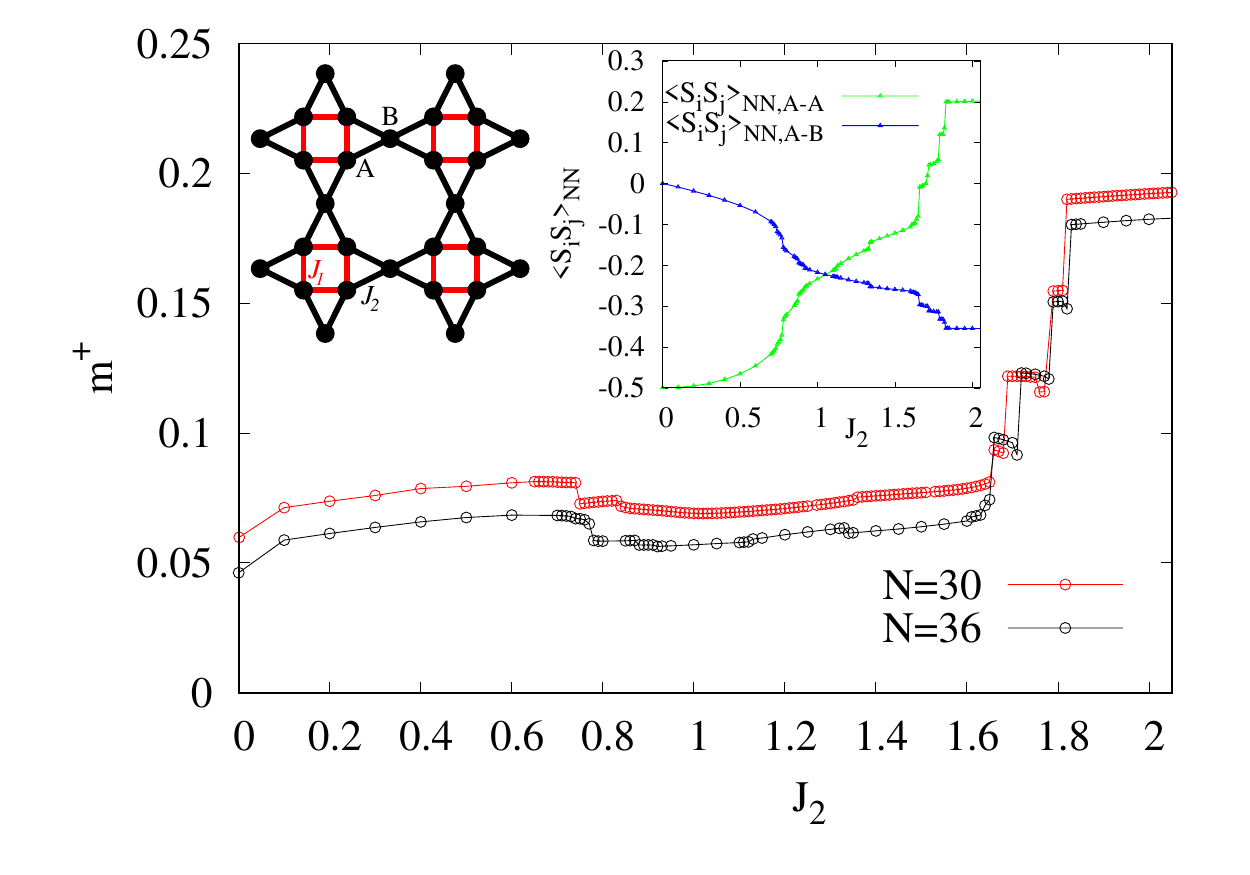}
\caption{Main panel: Order parameter $m^+$ as defined in Eq.~(\ref{mdef})  
of the spin-$1/2$ $J_1$-$J_2$ SKHAF ($N=30$ and $36$) as a function of $J_2$. 
Left inset: Sketch of the square-kagome lattice. Here A and B label the two non-equivalent sites
and $J_1$ and $J_2$ label the two non-equivalent nearest-neighbor bonds. 
Right inset: Nearest-neighbor spin-spin correlation for $N=36$:
$\langle{\bf s}_{i}\cdot{\bf s}_{j}\rangle_{NN,A-A}=
(\langle{\bf s}_{0}\cdot{\bf s}_{1}\rangle+\langle{\bf s}_{0}\cdot{\bf s}_{3}\rangle/2$
and $\langle{\bf s}_{i}\cdot{\bf s}_{j}\rangle_{NN,A-B}=
(\langle{\bf s}_{0}\cdot{\bf s}_{4}\rangle+\langle{\bf s}_{1}\cdot{\bf s}_{4}\rangle/2$, see
Fig.~\ref{fig_lat} for the numbering of sites. 
 }
\label{fig_m+}
\end{figure}

To get an impression on possible ground state magnetic order we first consider
an order
parameter introduced in Ref.~\cite{LNP645} that measures the total strength of the overall
spin-spin correlations without any assumptions on possible magnetic order 
with a related ordering vector $\bf Q$. It is defined as
\begin{equation} \label{mdef}
  m^+     =\frac{1}{N^2}
       \sum_{i,j}^{N}  {}|\langle{\bf s}_{i}\cdot{\bf s}_{j}\rangle|.
\end{equation}
Numerical ground-state data for $m^+$ are depicted in
Fig.~\ref{fig_m+}, main panel. It is obvious that in a wide parameter range
$0 \le J_2 \lesssim 1.65$  the order parameter $m^+$ is approximately of
the same small size as for the balanced model ($J_2=1$) which is known to be
in a non-magnetic singlet ground state.
Thus we may argue that there is no magnetic ground-state state order for $J_2 \lesssim
1.65$.
The steep increase of  $m^+$ beyond  $J_2 \approx 1.65$ is related to a
transition from a singlet ground state to a ferrimagnetic ground state with
non-zero magnetization $M$.  The jumps in the $m^+(J_2)$ curve visible for
$1.65 \lesssim J_2 \lesssim 1.83$  are related to a stepwise increase of $M$ up to $M_{1/3}=M_{\rm sat}/3$.
The ground state with $M_{1/3}$ present for $J_2 \gtrsim 1.83$ is a
ferrimagnetic  up-up-down (UUD) state, i.e., 
$\langle s^z_{i \in A}\rangle $ and $\langle s^z_{i \in B}\rangle$ are
antiparallel.
To give an example, for $N=36$, $J_2=2$, we have 
$\langle s^z_{i \in A}\rangle =0.39779$ 
and $\langle s^z_{i \in B}\rangle=-0.29558$.
We mention that for the classical model the 
transition to the UUD state
takes place  at $J_2=2$, i.e., the order-by-disorder mechanism
\cite{villain,shender1} leads to
a shift of the transition to the collinear UUD state to smaller values
of $J_2$.  
Bearing in mind  the  Schwinger-boson mean-field study of the ground state 
reporting 5 ground state phases \cite{Lugan2019}
it is worth to have a closer look on the details of the $m^+(J_2)$ profile.
Indeed, there are small discontinuous changes in  $m^+$ at $J_2 \approx 0.77$, $J_2 \approx 0.87$
and  $J_2 \approx 1.33$  
($J_2 \approx 0.74$, $J_2 \approx 0.83$ and  $J_2 \approx 1.35$) for $N=36$
($N=30$), where the values at about $0.85$ and $1.33$ are close to transition
points reported in  \cite{Lugan2019}.
We also mention that below $J_2 \approx 0.77$ the spins on B-sites become
weakly coupled to the neighboring A-site spins, whereas the nearest-neighbor
correlations on the $J_1$ bonds asymptotically approach  the value of the square-plaquette
singlet ground state (see the right inset in Fig.~\ref{fig_m+}), i.e., the
system enters a plaquette  
ground-state phase at low values of  $J_2$.

\begin{figure}[ht!]
\centering
\includegraphics*[clip,width=1.2\columnwidth]{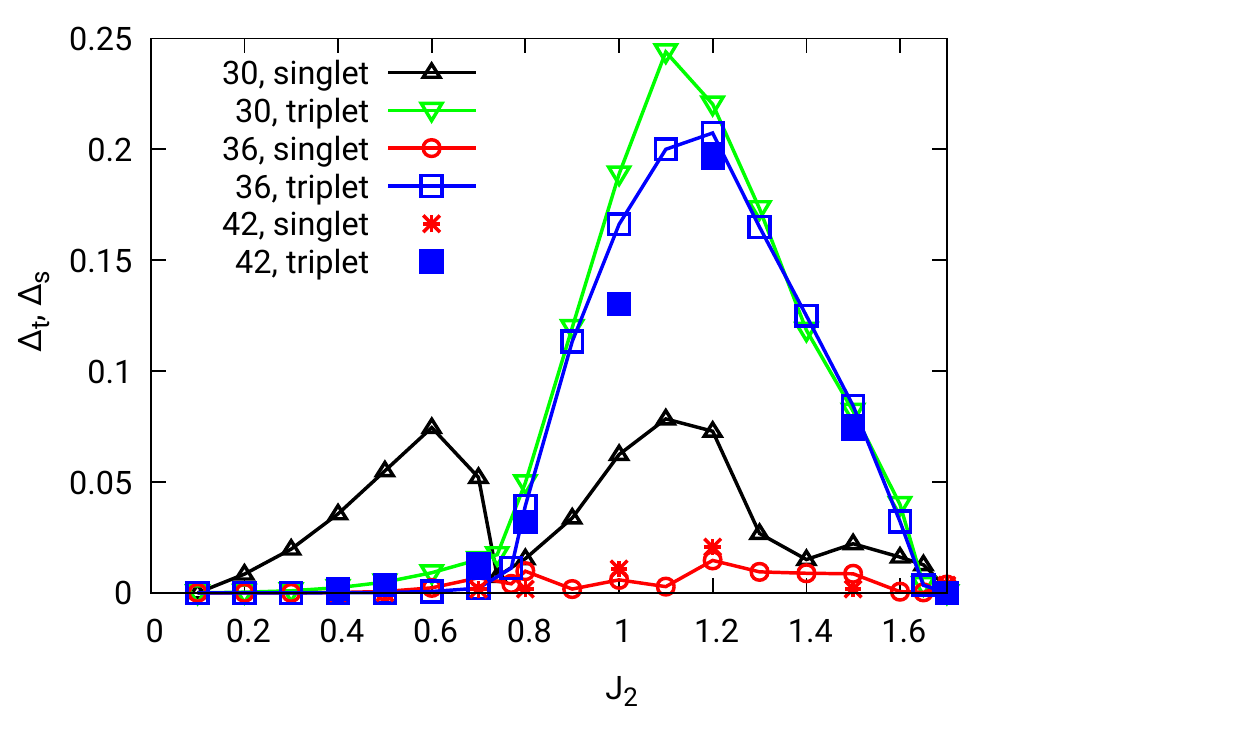}
\caption{Singlet-triplet and singlet-singlet gaps $\Delta_t$ and $\Delta_s$
of the spin-$1/2$ $J_1$-$J_2$ SKHAF ($N=30$ and $36$) as a function of $J_2$. 
For a few values of $J_2$ data for $N=42$  are added.
 }
\label{fig_gaps}
\end{figure}

For the low-temperature thermodynamics the spin gap (singlet-triplet gap)
$\Delta_t$ as well as the
singlet-singlet gap $\Delta_s$ are relevant.
Corresponding data are shown for $N=30$, $N=36$ and
$N=42$ in Fig.~\ref{fig_gaps}.
Our data provide evidence that there is a finite spin gap $\Delta_t$ in the region
between $J_2 \approx  0.77$ ($J_2 \approx  0.74$) and  $J_2 \approx  1.65$ ($J_2 \approx 1.65$)
for $N=36$ ($N=30$). We notice only a small finite-size dependence of the
spin gap away from $J_2=1$, 
whereas in the vicinity of $J_2=1$ it shrinks with
increasing $N$. However, it is known
that $\Delta_t$ remains finite at $J_2=1$ for  $N \to \infty$
\cite{schmoll2022tensor}. 
The vanishing of the spin gap at $J_2 \approx 0.77$ coincides with 
the above reported value at which a small discontinuous change in  $m^+$ 
occurs, whereas the closing of the spin gap at  $J_2 \approx 1.65$ is
related to the emergence of a ferrimagnetic ground state.
Thus,  $m^+$ as well as $\Delta_t$ yield indications for
ground-state phase transitions between a gapped and a gapless phase.
A similar behavior was found in Ref.~\cite{Lugan2019}, where, however, the
region of the gapped
phase is $0.84 \le J_2 \le 1.27$.
While $\Delta_t$ determines the low-temperature behavior of  the
susceptibility $X$,
the existence of low-lying singlet excitations within the spin gap, i.e., 
$\Delta_s < \Delta_t$, is crucial for the low-temperature behavior of  the
specific heat $C$.
From Fig.~\ref{fig_gaps}
it is obvious 
that in the whole region with a finite spin gap we have $\Delta_s <
\Delta_t$.  
As for the balanced model $J_1=J_2=1$ there are a number of singlets within the
spin gap. The details of their energy distribution will determine the
temperature profile of $C$ at very low $T$.      

\subsection{Thermodynamic properties of the SKHAF on finite lattices of
$N=30$ and $N=36$ sites}
\label{sec-3b}

Let us now consider the finite-temperature properties of the model. 
In what follows we discuss the Wilson ratio $P$,  the specific heat $C$, the
entropy $S$, and the uniform susceptibility $X$.

The modified Wilson ratio is defined as \cite{PRR2020_Wilson,Wilson_PRB2020} 
\begin{equation} \label{W}
P(T)=4\pi^2T X /(3NS) \;.
\end{equation}
It is a measure of 
the ratio of the density of magnetic excitations with $M>0$ 
and the density of {\it all} excitations  including singlet  excitations with
$M=0$.

\begin{figure}[ht!]
\centering
\includegraphics*[clip,width=0.95\columnwidth]{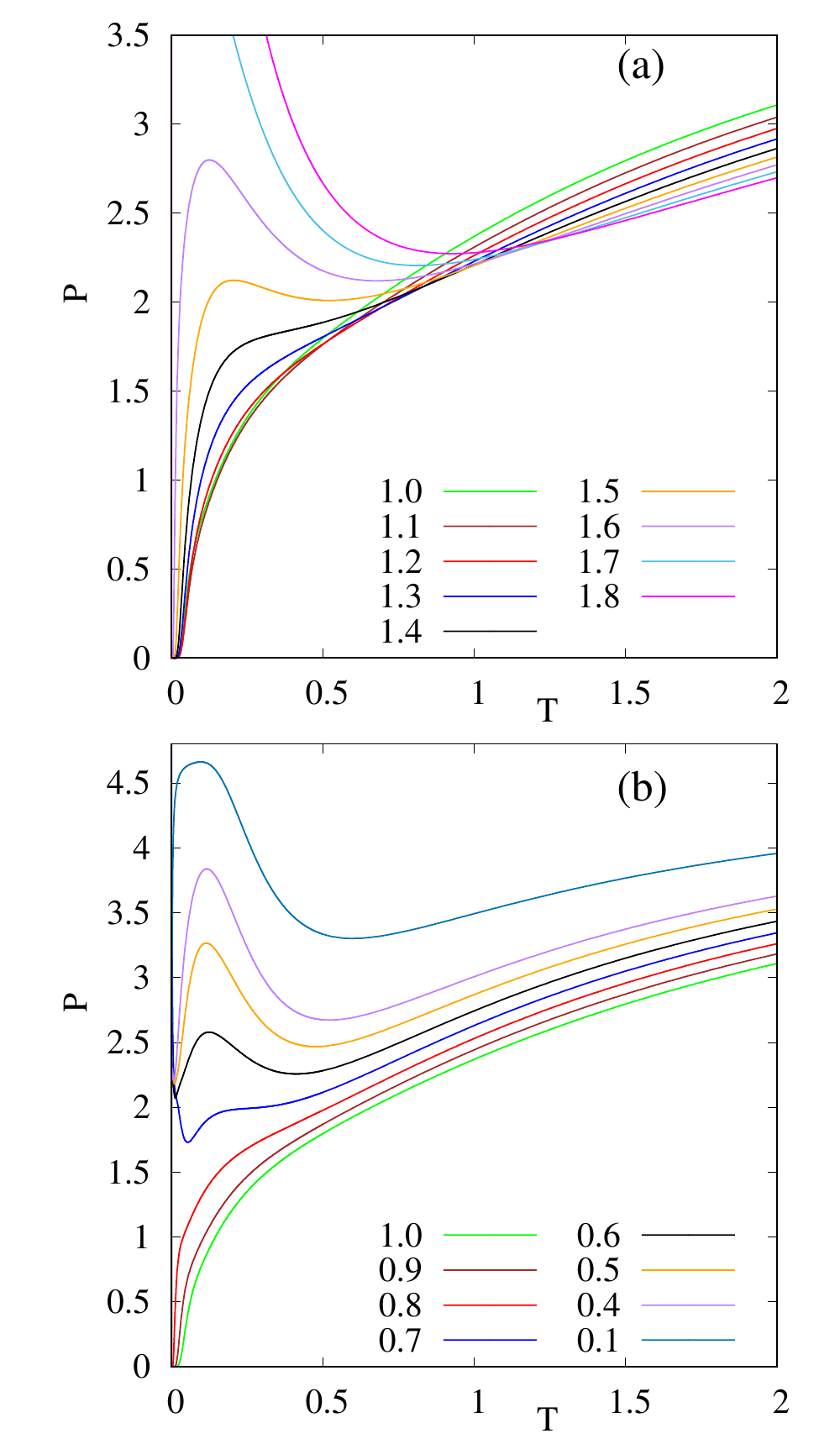}
\caption{Modified Wilson ratio $P(T)$, cf. Eq.~(\ref{W}),
of the spin-$1/2$ $J_1$-$J_2$ SKHAF ($N=36$). 
(a)  $J_2 \ge 1.0$, (b) $J_2 \le 1.0$.
 }
\label{fig_W}
\end{figure}

As shown for the KHAF \cite{PRR2020_Wilson,Wilson_PRB2020}  and for the
balanced SKHAF  \cite{squago_TD_2022} 
a vanishing $P$  as temperature $T \to 0$   
is a hallmark of a quantum spin-liquid ground state with dominating singlet
excitations at low $T$.
In contrast, for quantum spin models with semi-classical magnetic ground-state
order, 
such as the square-lattice Heisenberg antiferromagnet, the  Wilson ratio diverges
according to a power-law
\cite{PRR2020_Wilson,Wilson_PRB2020}. 
We show the  modified Wilson ratio in Fig.~\ref{fig_W}.
For $J_2=0.8,0.9,1.0,1.1,1.2,,1.3,1.4,1.5$ singlet excitations are
noticeably below the first triplet excitation.
As a result there is an obvious  downturn of $P$ as $T \to 0$.
Also the upturn of $P$ as $T \to 0$ for $J_2=1.7$ and  $1.8$ (ferrimagnetic ground
state) is evident.
More subtle is the situation for $J_2<0.8$, where 
the plaquette ground-state phase emerges. Here the low-lying spectrum is
dominated by the weakly coupled spins on the B sites.
which leads to a maximum in $P$ at low temperatures, see
Fig.~\ref{fig_W}(b).
This behavior can be understood by considering  the ground state in the limit of decoupled B
spins, i.e., for $J_2=0$. In this limit we get a size independent Wilson
ratio $P_0=\lim_{T \to 0}P=\pi^2/(3\ln2)=4.74628$. Obviously, the height of the
low-temperature maximum in $P$ approaches $P_0$ as decreasing $J_2$.
At very low $T$ the Wilson
ratio approaches a constant value of about $P \approx 2$. (Note, however,
that our FTLM is not appropriate to get accurate data precisely at $T=0$,
because in the limit of very weakly coupled B spins very tiny energy
differences appear in the low-energy spectrum.)      
As reported in Ref.~\cite{PRR2020_Wilson}  this behavior corresponds to a
gapless spin liquid; in particular, for the one-dimensional $s=1/2$ Heisenberg
antiferromagnet (Bethe chain) $P_0$ is
exactly $2$ \cite{PRR2020_Wilson,PhysRevB.61.9558}.

\begin{figure}[ht!]
\centering
\includegraphics*[clip,width=1.0\columnwidth]{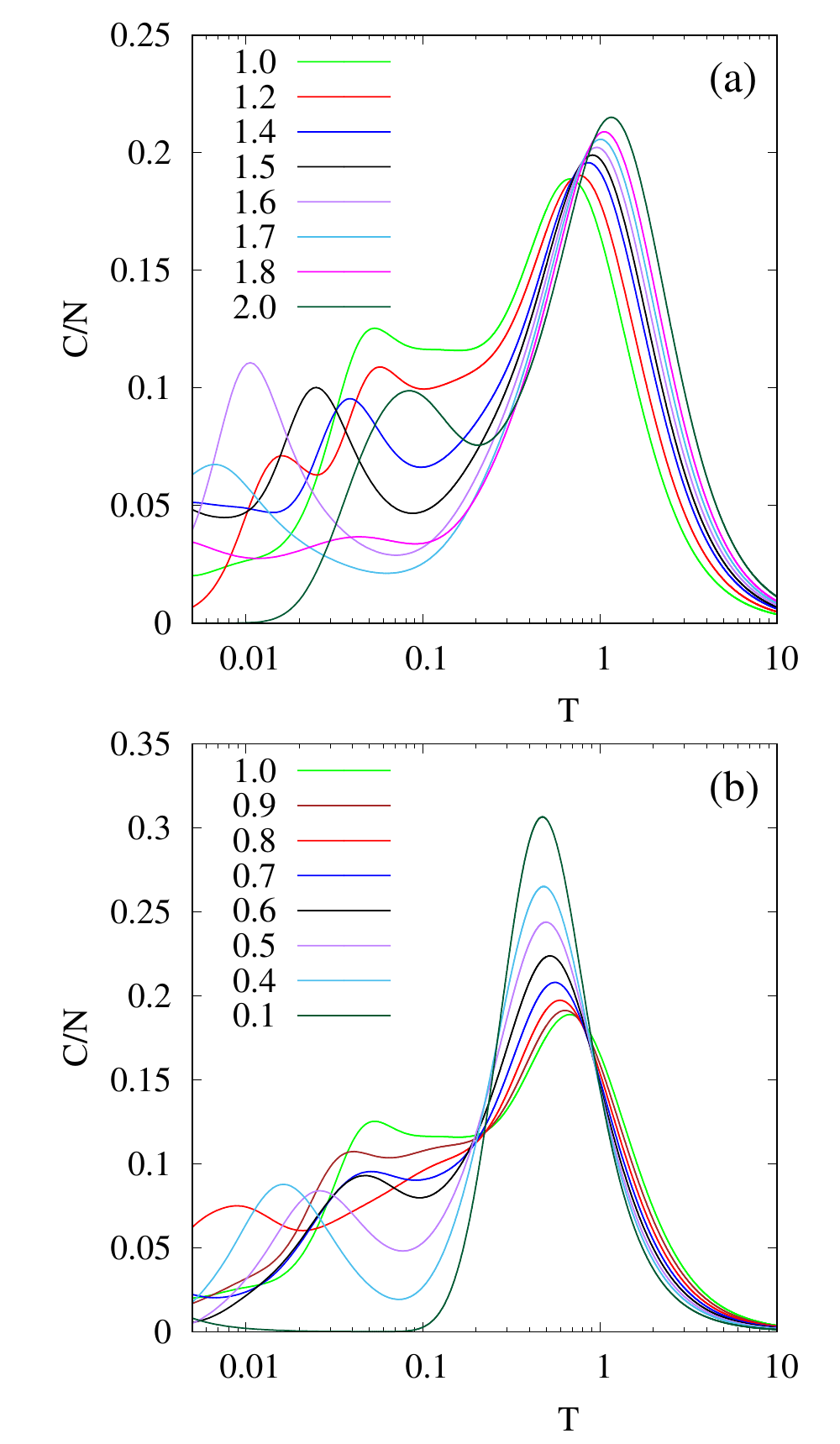}
\caption{Specific heat $C/N$ of the spin-$1/2$ $J_1$-$J_2$ SKHAF for $N=36$. (a)  $J_2 \ge 1.0$,
(b) $J_2 \le 1.0$.
}
\label{fig_C}
\end{figure}

\begin{figure}[ht!]
\centering
\includegraphics*[clip,width=1.05\columnwidth]{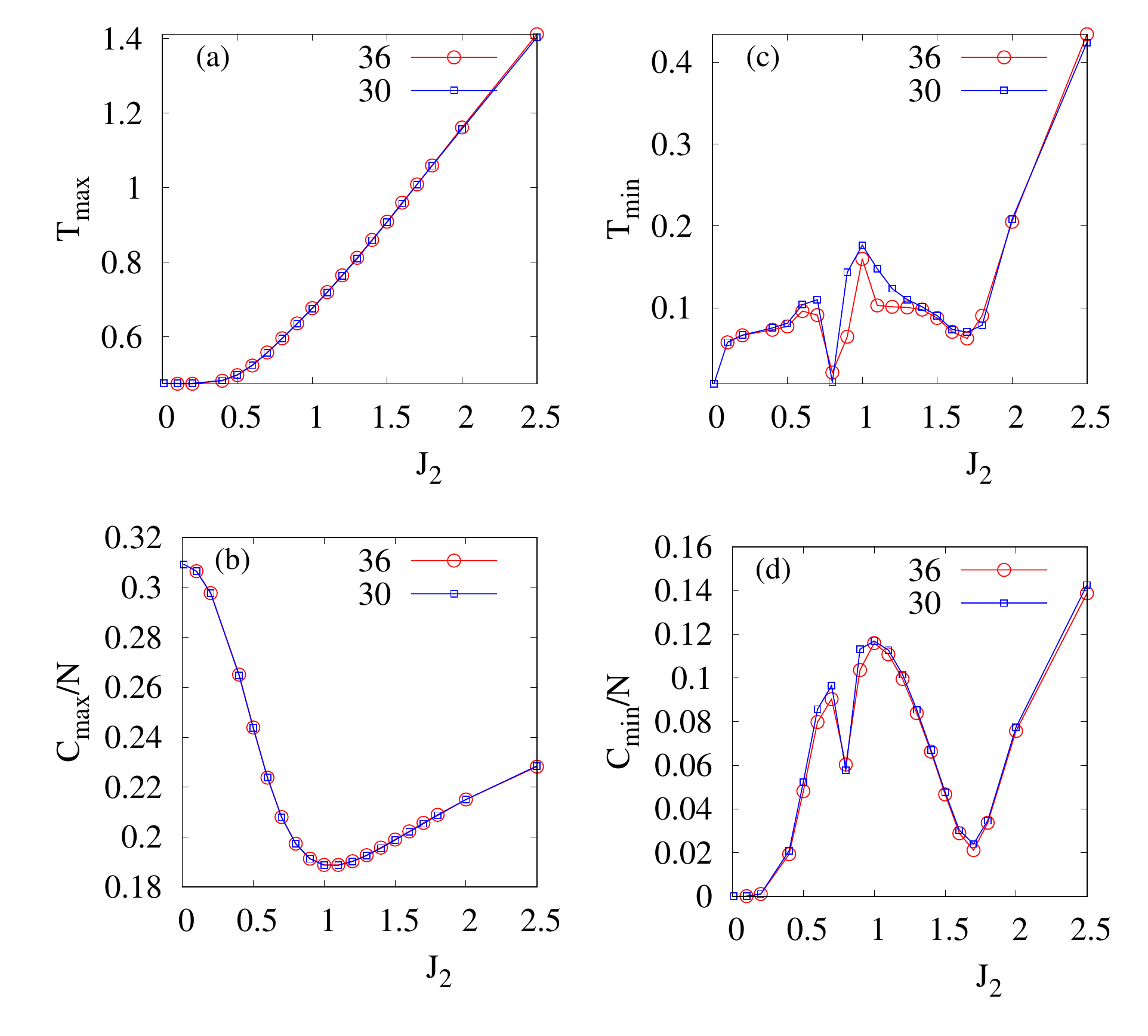}
\caption{Features of the main maximum and the minimum below the main maximum 
in the temperature profile of the specific heat $C(T)/N$ of the spin-$1/2$ $J_1$-$J_2$ SKHAF
($N=30$ and $N=36$).  
(a) Position $T_{\rm max}$ and (b) height $C_{\rm max}/N$ of the main
maximum.
(c) Position $T_{\rm min}$ and (d) depth $C_{\rm min}/N$ of the 
minimum.
}
\label{fig_Cmax_Tmax}
\end{figure}

\begin{figure}[ht!]
\centering
\includegraphics*[clip,width=1.0\columnwidth]{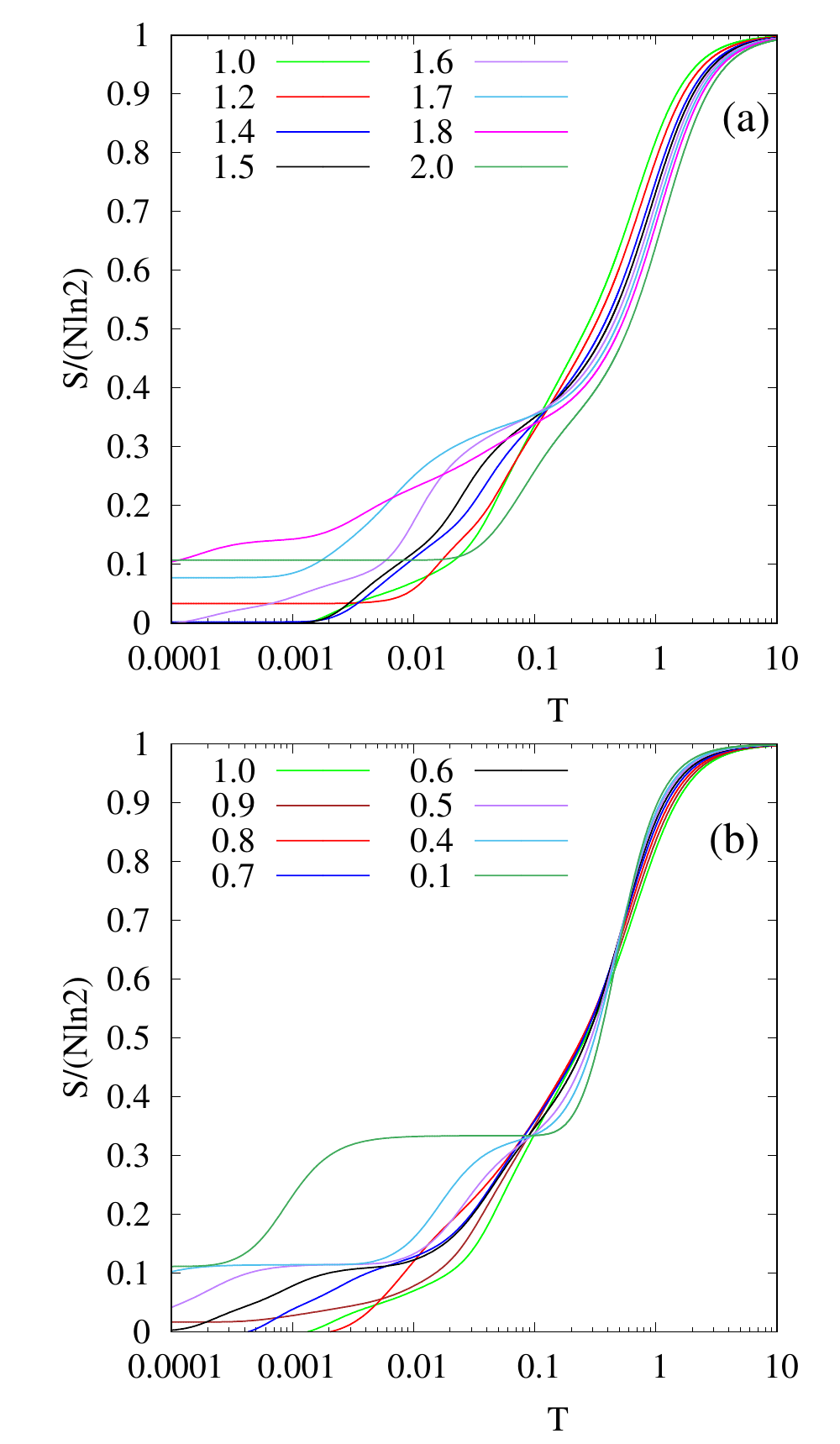}
\caption{
Entropy $S/N$ of the spin-$1/2$ $J_1$-$J_2$ SKHAF for $N=36$. (a)  $J_2 \ge 1.0$,
(b) $J_2 \le 1.0$. 
Note that the finite entropy at $T=0$ for  $J_2=1.8$ and
$2.0$ is caused by the 
ferrimagnetic multiplet and by an accidental degeneracy of the
ground state for some other values of $J_2$.
}
\label{fig_S}
\end{figure}

\begin{figure}[ht!]
\centering
\includegraphics*[clip,width=1.0\columnwidth]{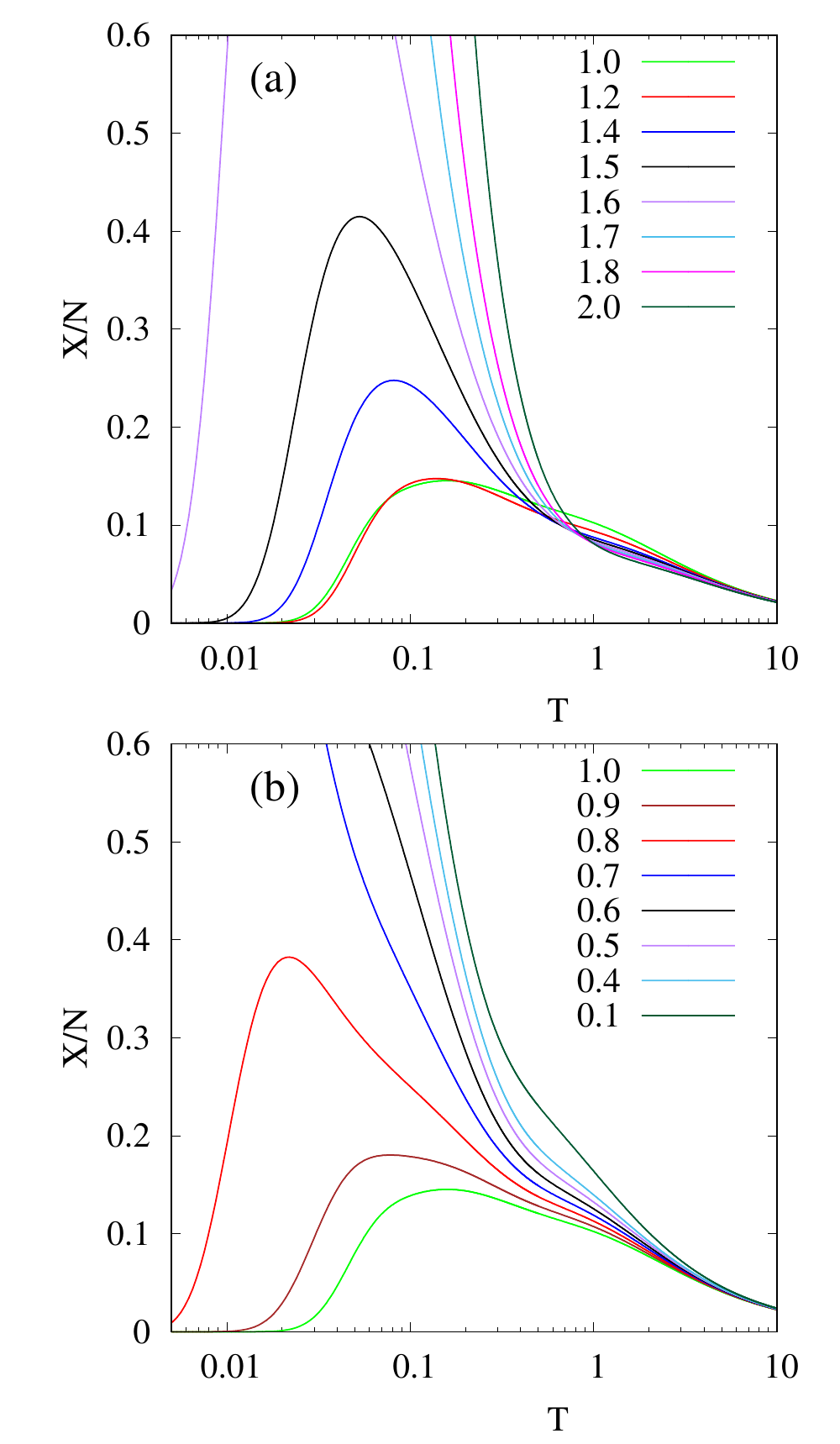}
\caption{Susceptibility  $X/N$ of the spin-$1/2$ $J_1$-$J_2$ SKHAF for $N=36$. (a)  $J_2 \ge 1.0$,
(b) $J_2 \le 1.0$.
}
\label{fig_chi}
\end{figure}

\begin{figure}[ht!]
\centering
\includegraphics*[clip,width=1.0\columnwidth]{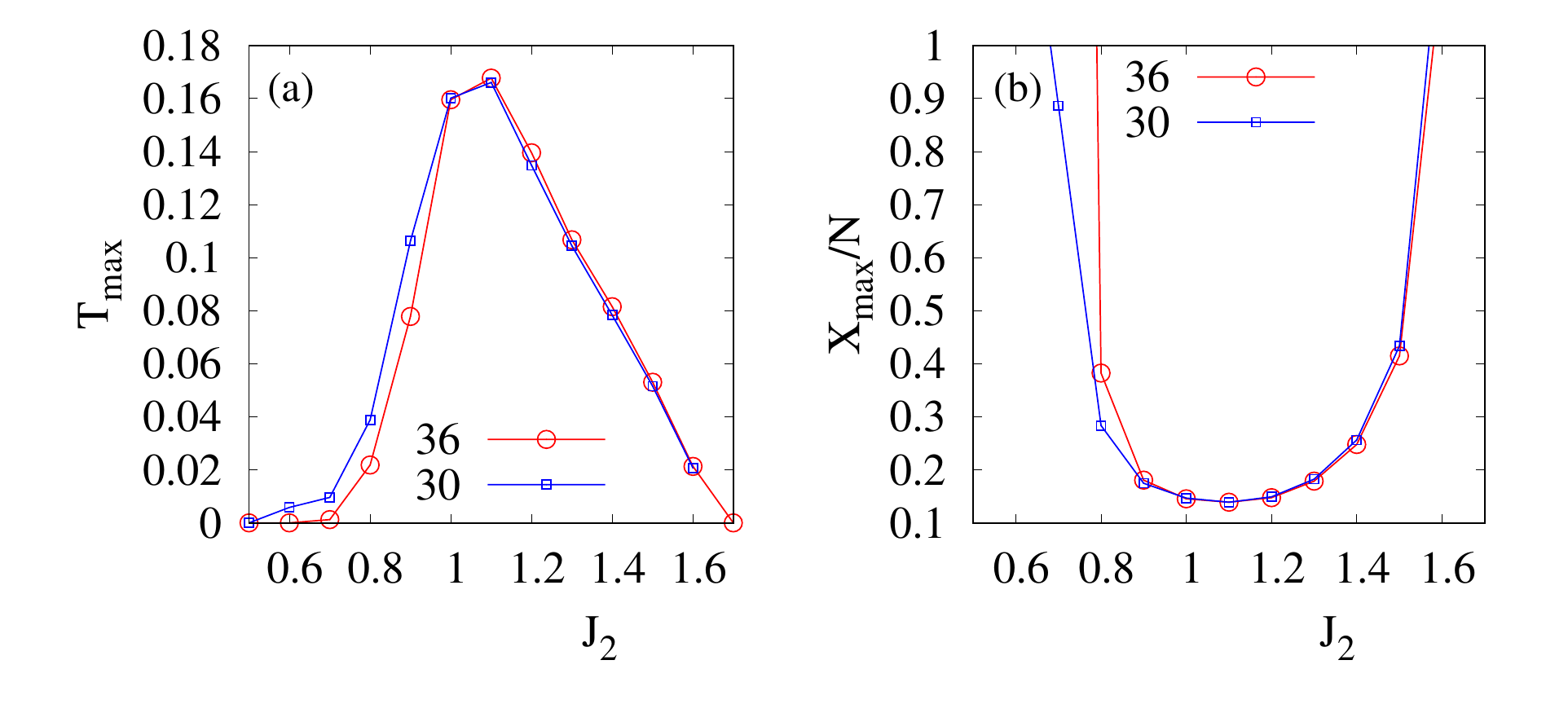}
\caption{Features of the maximum 
in the temperature profile of the susceptibility  $X(T)/N$ of the spin-$1/2$ $J_1$-$J_2$ SKHAF
($N=30$ and $N=36$).  
(a) Position $T_{\rm max}$ of the maximum.
(b) Height $X_{\rm max}/N$ of the maximum.
}
\label{fig_Tchimax}
\end{figure}

Let us now discuss the specific heat $C(T)$, the
entropy $S(T)$ and the uniform 
susceptibility $X(T)$. We use a logarithmic temperature scale which
makes the 
low-temperature features transparent, see
Figs.~\ref{fig_C}, \ref{fig_S}, and \ref{fig_chi}.
In panels (a) we show data for $J_2 \ge 1$ and in  panels (b) for $J_2 \le
1$. 
In all these figures we also show the corresponding data for the balanced model
\cite{squago_TD_2022}
which may serve as benchmark data. 
The typical main maximum is related to the magnitudes of $J_1$ and $J_2$. Its
position $T_{\rm max}$ and its height $C_{\rm max}/N$ exhibit a quite regular
behavior, see Fig.~\ref{fig_Cmax_Tmax} (a) and (b).

From  Fig.~\ref{fig_Cmax_Tmax} it is also evident that $T_{\rm max}$ and $C_{\rm
max}/N$ are
equal for $N=30$ and $36$ for all values of $J_2$, i.e., the main maximum in $C(T)/N$ is  
not affected by finite-size effects, see also Fig.~\ref{fig_C_app} in Appendix 
\ref{sec-a2}.
For all values of $J_2$ shown in Fig.~\ref{fig_C} the temperature profile exhibits 
a low-temperature maximum 
below the main maximum that indicates an extra-low energy scale. 
Though, we show in Fig.~\ref{fig_C} only data for $N=36$  this feature is present
also for $N=30$ and $N=42$, cf.  Fig.~\ref{fig_C_app} in Appendix \ref{sec-a2}.
Since the $C(T)$ curves for $N=30,36,42$ coincide down to
temperatures where this particular low-$T$ feature emerges, 
we may argue that this characteristic survives for $N \to \infty$ either as
an extra maximum or a shoulder below the main maximum.   
An additional information on the finite-size dependence of the low-$T$  part
of $C(T)$ is given in Fig.~\ref{fig_Cmax_Tmax}, where we show the position 
$T_{\rm min}$ [panel (c)] and the depth $C_{\rm min}/N$    
of the minimum [panel (d)] in $C(T)$ below the main maximum. 
The good agreement of the data for $N=30$ and $36$ is obvious.
The special values of $T_{\rm min}$ and $C_{\rm min}/N$  found for $J_2=0.8$ might be
attributed to the proximity to the transition point to the plaquette
phase. 
Interestingly, there is also a double-maximum profile in $C(T)$ for
$J_2=1.8$ and $2.0$, where the ground state is ferrimagnetic. Only beyond $J_2
\sim 3$ we get  a $C(T)$ profile with only one maximum, see
Fig.~\ref{fig_C_large_J2_app} in 
Appendix \ref{sec-a2}.
Let us finally mention that  for some values of
$J_2$ there is even some additional structure at very low $T \lesssim 0.02$
which most likely can be attributed to finite-size effects.

In highly frustrated quantum magnets we may have a high density of states at
low excitation energies \cite{kago42,squago_TD_2022,krivnov2014delta}. 
To shed light on the density 
of low-lying  eigenstates 
we present  the entropy $S(T)/N$ 
in Fig.~\ref{fig_S}. We observe, that already at $T \sim 0.2$ about $50\%$ of the maximum
entropy $S(T\to \infty)=N\ln2$ is acquired. Note that for the unfrustrated
square-lattice Heisenberg antiferromagnetthe corresponding value at  $T \sim 0.2$ is only about
$10\%$, cf. Ref.~\cite{kago42}.
Moreover, there is a  
 change in the curvature or even a  plateau-like feature  in the $S(T)$ profile
below this temperature. In particular, for $J_2 \lesssim 0.7$ we see such a
plateau at  $S/(N\ln2) \sim 0.1$ which can be attributed to a high density of
states  caused by the weakly coupled B spins in this parameter region.
For some values of $J_2$ (e.g. for $J_2=2.0$ and $1.2$) there is a
finite value of $S(T=0)/N$ due a degeneracy of the ground state.
However, $S(T=0)/N$ will become zero as $N \to \infty$.
For more information on finite-size effects, see Fig.~\ref{fig_S_app} in
Appendix \ref{sec-a2},
where data for $N=30$, $N=36$  and $N=42$ are compared.

Next we turn to the zero-field  susceptibility $X$ displayed
in Fig.~\ref{fig_chi}.  
For $J_2$ values where we have a finite singlet-triplet gap $\Delta_t$, see Fig.~\ref{fig_gaps},
 $X$ exhibits
 an exponentially activated low-temperature behavior and there is a maximum
in $X(T)$.
Its position $T_{\rm max}$ and its height $X_{\rm max}/N$ exhibit a quite regular
behavior, and the finite-size effects in 
$T_{\rm max}$ and $X_{\rm max}/N$ are small, see Fig.~\ref{fig_Tchimax}.
(For more information on finite-size effects, see Fig.~\ref{fig_chi_app} in Appendix \ref{sec-a2},
where data for $N=30$, $N=36$, and $N=42$ are compared.)
Around $J_2=1$ the position $T_{\rm max}$ is largest, although,
it is still  at a pretty low temperature
compared to  $T_{\rm max}=0.935$  for the
square-lattice Heisenberg antiferromagnet\cite{Johnston2011,SLR:PRB11}, which demonstrates 
the crucial role of frustration also for the susceptibility.   
It is also obvious that  $T_{\rm max}$ is directly related to the spin gap
$\Delta_T$, compare Fig.~\ref{fig_Tchimax}(a) and Fig.~\ref{fig_gaps}.
Increasing $J_2$ towards the transition to the ferrimagnetic ground state
naturally leads to a diminishing of $T_{\rm max}$  and an increase of  $X_{\rm max}/N$.  
At $J_2 \sim 1.65$ we get $T_{\rm max}=0$ and  
$X_{\rm max}/N \to \infty$.

A similar behavior can be observed for decreasing $J_2$ towards $J_2=0$.
Again the  singlet-triplet gap $\Delta_t$ becomes smaller and it is
effectively zero below $J_2 \sim 0.77$ (plaquette  
ground-state phase), i.e., $T_{\rm max}$ approaches zero.
However, here the weakly coupled spins on the B sites lead to extremely
low-lying magnetic and non-magnetic excitations.
In fact, we find that for the finite systems considered here the ground
state is still a non-magnetic singlet but magnetic excitations dominate
the $X(T)/N$ profile down to very low $T$.
Thus, the susceptibility indeed vanishes at $T=0$, but $X(T)/N$  approaches
zero only at $T \sim 10^{-4}$, $10^{-5}$, 
 $10^{-9}$ for $J_2=0.7$, $0.5$, $0.1$, respectively.

\subsection{Field-dependent properties}
\label{sec-3c}

\begin{figure}[ht!]
\centering
\includegraphics*[clip,width=1.02\columnwidth]{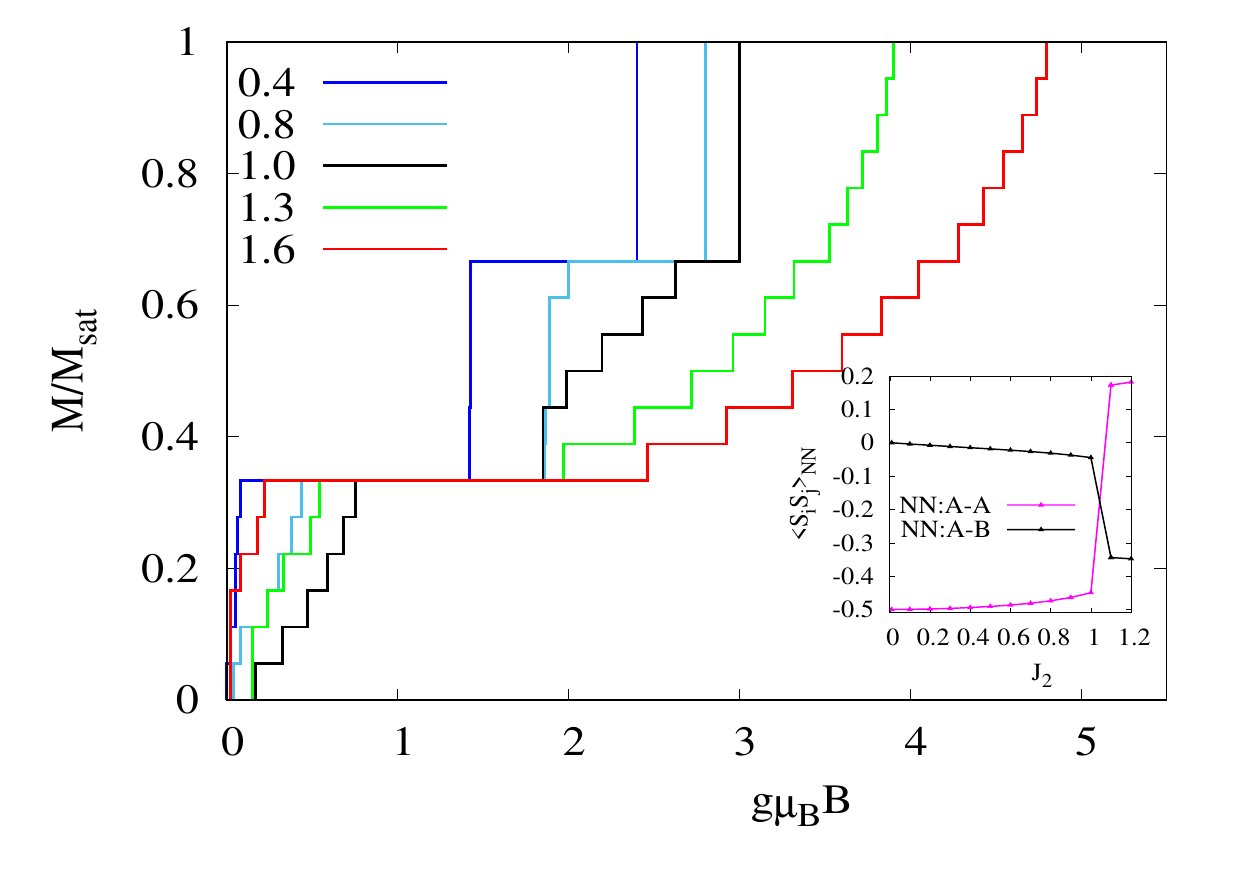}
\caption{Main panel: Zero-temperature magnetization curves of the $J_1$-$J_2$ SKHAF with $N=36$
sites and selected values of $J_2$. Inset:
Nearest-neighbor correlation functions
$\langle{\bf s}_{i}\cdot{\bf s}_{j}\rangle_{NN,A-A}$ and $\langle{\bf s}_{i}\cdot{\bf
s}_{j}\rangle_{NN,A-B}$ in the $1/3$ plateau state.
 }
\label{fig_M_H}
\end{figure}

\begin{figure}[ht!]
\centering
\hspace*{-0.7cm}\includegraphics*[clip,width=1.1\columnwidth]{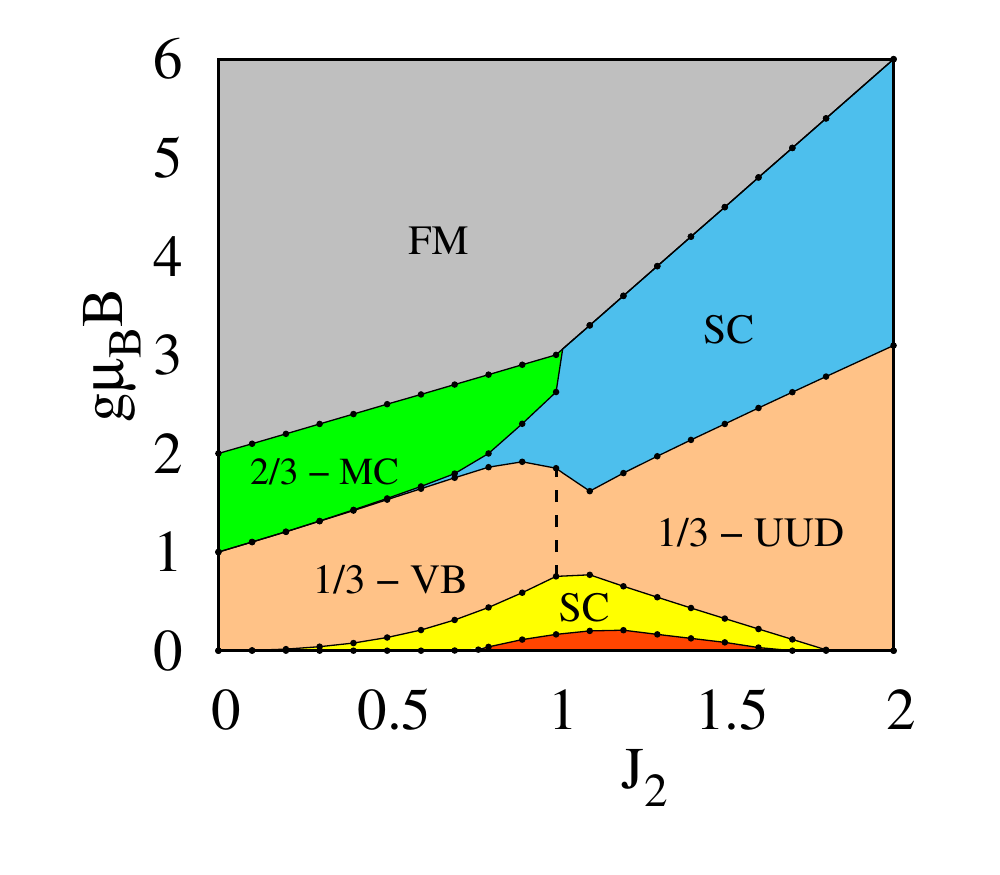}
\caption{Sketch of the zero-temperature $J_2$-$B$ phase
diagram. FM: ferromagnetic, 1/3-VB: $M=M_{\rm sat}/3$ (valence-bond state),
1/3-UUD: $M=M_{\rm sat}/3$ (up-up-down state), 2/3-MC: $M=2M_{\rm sat}/3$
(magnon crystal), SC: spin canting above and below the 1/3
plateau, red area: $M=0$ (gapped phase).
 }
\label{fig_PD}
\end{figure}

The magnetization process of strongly frustrated quantum magnets exhibits   
a number of
interesting features, such as plateaus and jumps \cite{HSR:JP04}.
Previous studies for the balanced model
\cite{richter2009squago,Sakai2013,squago_TD_2022,asym_melting_2022,schmoll2022tensor}
report on wide  plateaus at $1/3$ and $2/3$ of the
saturation magnetization $M_{\rm sat}$.
Moreover, there is the typical macroscopic
jump to saturation due to the presence of independent localized
multi-magnon ground states stemming from a flat one-magnon
band \cite{SSR:EPJB01,SHS:PRL02,zhitomirsky2004exact,DRH:LTP07}.

Let us first present the zero-temperature magnetization curve for selected
values of $J_2$, see Fig.~\ref{fig_M_H}. 
Both plateaus as well as the jump to saturation known from the balanced 
model are present for all values
$J_2\le 1$, whereas for $J_2 > 1 $ the jump and the related preceding 
$2/3$ plateau are missing. 
For $J_2\le 1$ both plateau states are
non-classical valence-bond states, cf.
Refs.~\cite{richter2009squago,squago_TD_2022,asym_melting_2022,schmoll2022tensor},
where the upper plateau state is the
exactly known magnon-crystal product state, i.e., spins on the $B$ sites are fully
polarized and the $A$-spins on a square occupy the lowest triplet eigenstate 
of the square plaquette with $S^z_{\rm plaqu}=1$, for an illustration of
this state, see, e.g., Fig.~2a in
Ref.~\cite{Derzhko2006universal}. For $J_2\lesssim 0.7$ the transition
between the two plateaus becomes steplike. The jump to saturation as well as the magnon-crystal
product state are related to the
flat one-magnon band which is the lowest one  for $J_2 \le 1$. 
In contrast, for $J_2> 1$, the flat one-magnon band is not the lowest
one, and, therefore the flat-band related features are not present in the
magnetization curve.
However, the flat-band related localized 
multi-magnon states including the magnon crystal
are still eigenstates living now as quantum scar states somewhere in the
middle of the spectrum \cite{McClarty2020}.

The valence-bond state of the lower plateau is not exactly known but it 
is approximately  described 
by a  product state with fully polarized spins on the $B$ sites and  a
singlet state of the $A$-spins on a square, see the inset in
Fig.~\ref{fig_M_H}, where the  spin-spin correlations 
$\langle{\bf s}_{i}\cdot{\bf s}_{j}\rangle_{NN,A-A}$ and $\langle{\bf s}_{i}\cdot{\bf
s}_{j}\rangle_{NN,A-B}$
in the $1/3$ plateau state are shown.
On the other hand, for  $J_2> 1$ the $1/3$ plateau state is
semi-classical, namely it is the ferrimagnetic UUD state, cf.
Sec.~\ref{sec-3a} and see the inset in
Fig.~\ref{fig_M_H}.

Using the entire set of calculated  magnetization curves for
$N=36$  (which includes  altogether  22 $J_2$
values in the  region $0 \le J_2 \le 2$)  we can construct the  $J_2$-$B$ phase
diagram shown in Fig.~\ref{fig_PD}. 
The saturation magnetization (uppermost line) is given 
by $g \mu_B\, B_{\rm sat}=2+J_2$
 for $J_2 \le 1$ and $g \mu_B\, B_{\rm sat}=3J_2$ for $J_2 \ge 1$, and it is
size-independent.
But also
for the other phase boundaries derived from numerical data, the finite-size effects are very
small, see Fig.~\ref{fig_width} in appendix \ref{sec-a2}.

For
elevated temperatures the experimental detection of plateaus may become
intricate, because often there is a fast melting of plateaus and jumps,
i.e.,
they are smeared out already at pretty low $T$,
see, e.g., Refs.~\cite{Misawa2018,asym_melting_2022}.
Therefore, to detect plateaus and jumps in experiments   
the differential susceptibility $X(T,B)=dM(T,B)/d(g \mu_B\, B)$ 
as a function of $B$ measured at various $T$ is more suitable, cf., e.g., Ref.~\cite{Tanaka_trian_2012}.
Magnetization plateaus  show up as pronounced minima  in
$X(B)$, however, requiring sufficiently low
temperatures. On the other hand,
a jump of the magnetization 
leads to a high peak in $X(B)$ at low $T$.

\begin{figure}[ht!]
\centering
\includegraphics*[clip,width=1.1\columnwidth]{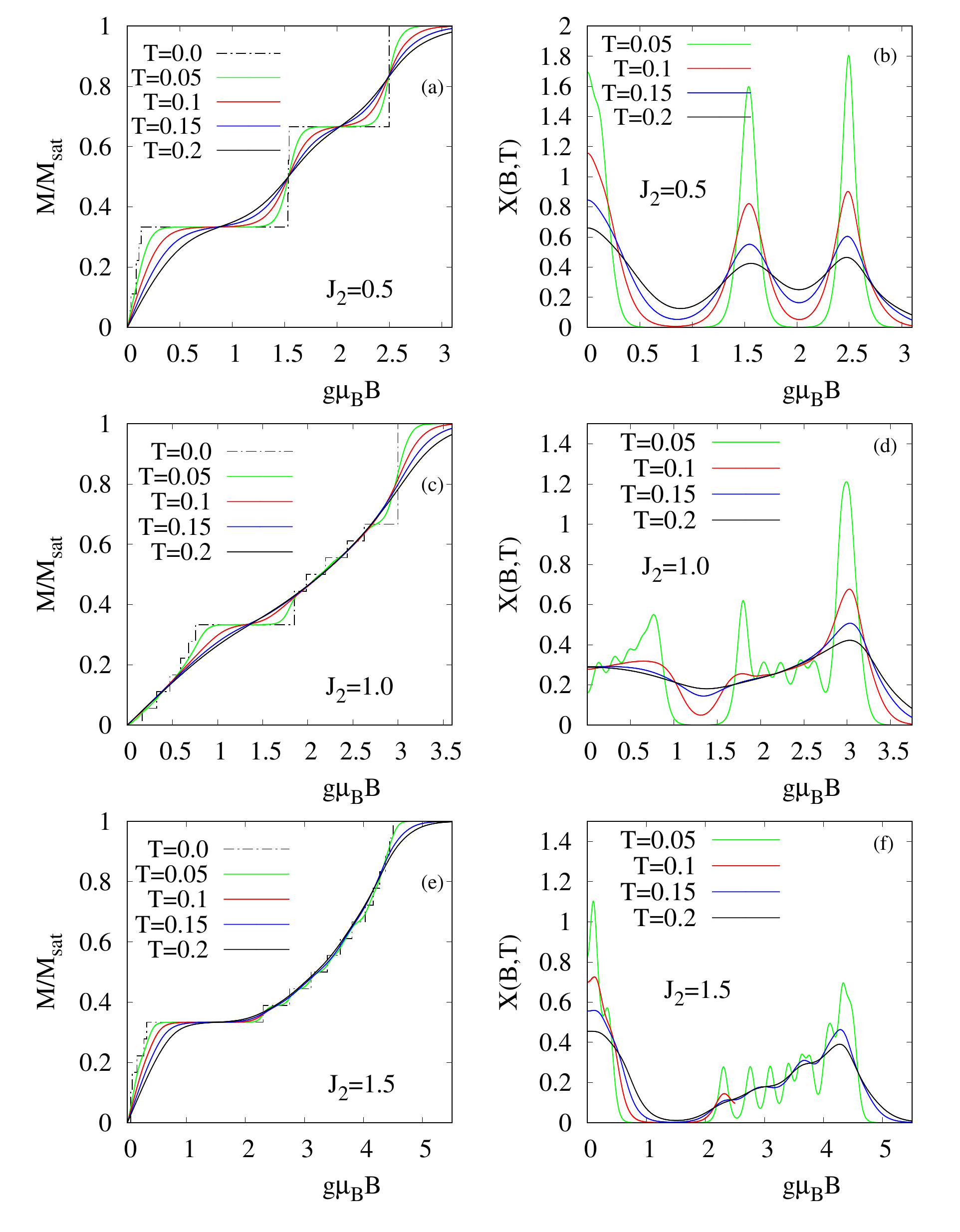}
\caption{ Left panels: Finite temperature magnetization curves $M(T,B)$ of the $J_1$-$J_2$ SKHAF with $N=36$
sites for selected values of $J_2$. 
Right panels: Corresponding data of the differential susceptibility 
$X(T,B)=dM(T,B)/d(g \mu_B\, B)$.
 }
\label{fig_M_H_T}
\end{figure}

We present the influence  of the temperature on the magnetization curve  $M(B)$
and on the differential susceptibility $X(B)$
in
Fig.~\ref{fig_M_H_T} for selected values of $J_2$.
We observe, that the melting process is most rapid for $J_2 \sim
1$, whereas for small and large $J_2$ the plateaus and the jumps are still well visible at
$T=0.2$.
We notice that the oscillations present for $J_2 \ge 1$  at $T=0.05$ (green
curves) above the $1/3$ plateau
are finite-size effects.

\section{Summary and Conclusions}
\label{sec-4}

In our study
 we performed  numerical calculations of
thermodynamic quantities such as the magnetization $M(T)$, the specific
heat $C(T)$, the entropy $S(T)$ and the susceptibility $X(T)$ for the $J_1$-$J_2$ spin-half square-kagome 
Heisenberg antiferromagnet
(SKHAF) by 
using the finite-temperature Lanczos method (FTLM) applied to
finite lattices of $N=30$, $N=36$  and $N=42$ sites.
Since the  
SKHAF exhibits two non-equivalent nearest-neighbor bonds,
the extension of previous studies \cite{tomczak2003specific,squago_TD_2022},
which were restricted to $J_1=J_2$, on the generalized model with  $J_1 \ne J_2$ is natural with respect to experimental realization
of the SKHAF, see Refs.~\cite{FMM:NC20,YSK:IC21,Vasiliev2022,Vasiliev2023}.
Moreover, the generalized model may serve as a model allowing to tune the
competition of antiferromagnetic bonds in a highly frustrated spin system.

The exact-diagonalization data for the ground state indicate magnetic
disorder in a wide range of $J_2/J_1$ ratios. Only for  $J_2 \gtrsim 1.65J_1$
the ground state features ferrimagnetic order.
In the region    $0.77 \lesssim J_2/J_1 \le 1.65$
the low-temperature thermodynamics is determined by a finite singlet-triplet
gap
with low-lying singlets within this gap.
Therefore, the susceptibility decays exponentially to zero as temperature
$T \to 0$, while the specific heat exhibits an extra maximum at low $T$ related
to the singlets.
For smaller values of  $J_2/J_1 \lesssim 0.7$ 
the ground state becomes  a plaquette  
ground state with weakly coupled spins on B sites which become asymptotically
decoupled as $J_2/J_1 \to 0$.
As a result, the entropy acquires already a large amount at very
low temperatures.

In non-zero magnetic field $B$ we find well pronounced plateaus at $1/3$ and $2/3$ of the saturation
magnetization and a jump from the $2/3$ plateau to saturation in the whole
region   $0 \le J_2/J_1 \le 1$, whereas for $J_2/J_1 > 1$ only the  $1/3$
plateau is present.
At  at low and moderate  temperature
the plateaus  are reflected as  minima in the
differential susceptibility $X(B)=dM(B)/d(g \mu_B\, B)$ and a jump is
seen as a peak in $X(B)$. 

Bearing in mind the numerous studies of the low-energy physics of the
related kagome Heisenberg antiferromagnet
we argue that our work may also stimulate other studies using
alternative techniques, such as tensor network
methods, DMRG, numerical linked cluster expansion or Green's function techniques
\cite{Pollmann2017,Xi-Chen2018,SHM:PRB20,kagome-RGM2018,Rausch2021,schmoll2022tensor,schmoll2022tensor_T}.

\section*{Acknowledgment}

This work was supported by the Deutsche
Forschungsgemeinschaft (DFG  RI 615/25-1 and SCHN 615/28-1). Computing time at
the Leibniz Center in Garching (project pr62to) is gratefully acknowledged.

\appendix
\section{Finite square-kagome lattices used for 
the exact diagonalization 
and the
finite-temperature Lanczos method}
\label{sec-a1}

\begin{figure}[ht!]
\includegraphics*[clip,width=1.0\columnwidth]{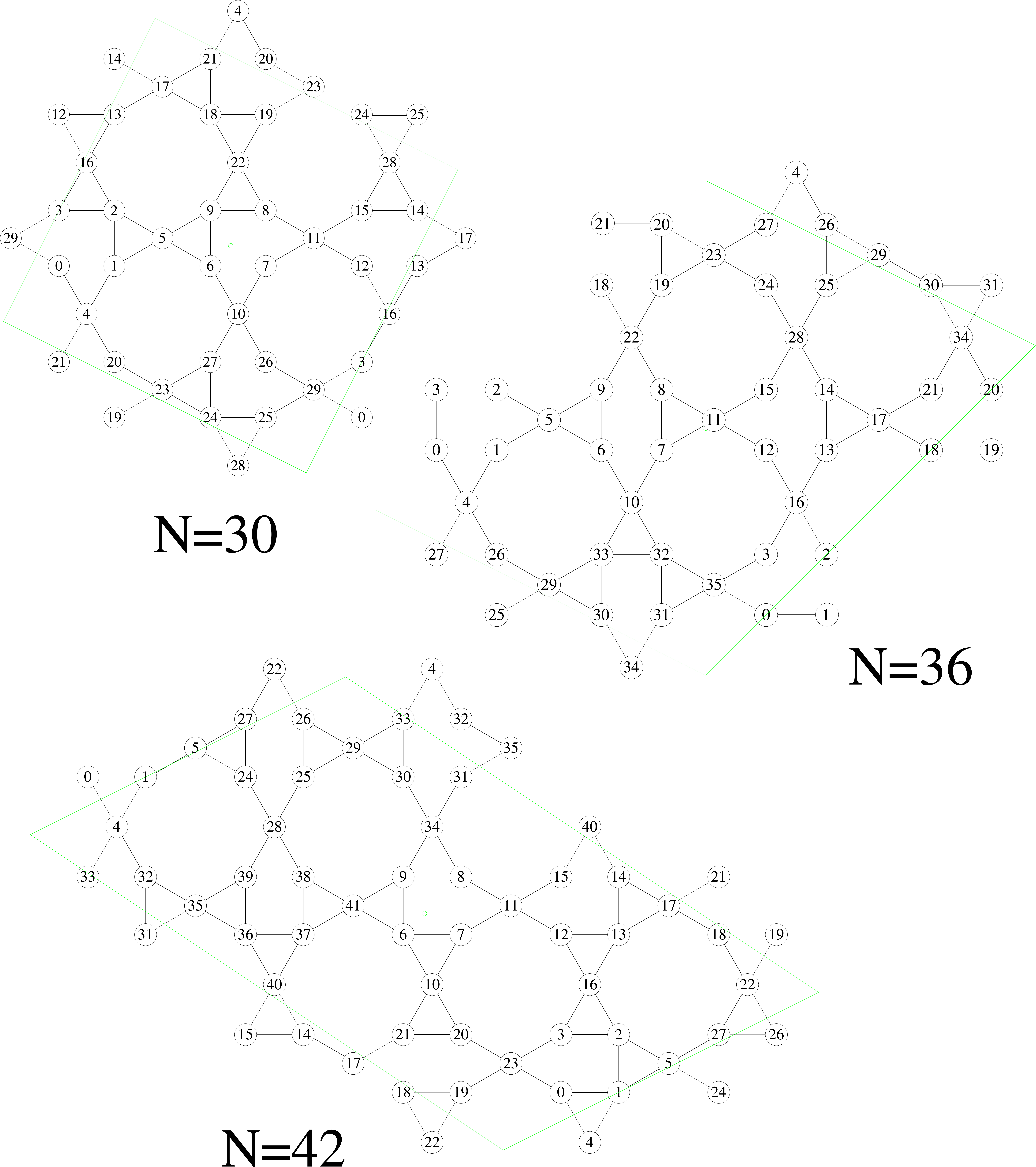}
\caption{Finite square-kagome lattices of $N=30$, $N=36$ and $N=42$ sites.
}
\label{fig_lat}
\end{figure}

Here we provide the finite lattices studied in our paper, see Fig.~\ref{fig_lat}.

\section{Finite-size effects}
\label{sec-a2}

\begin{figure}[ht!]
\centering
\includegraphics*[clip,width=1.2\columnwidth]{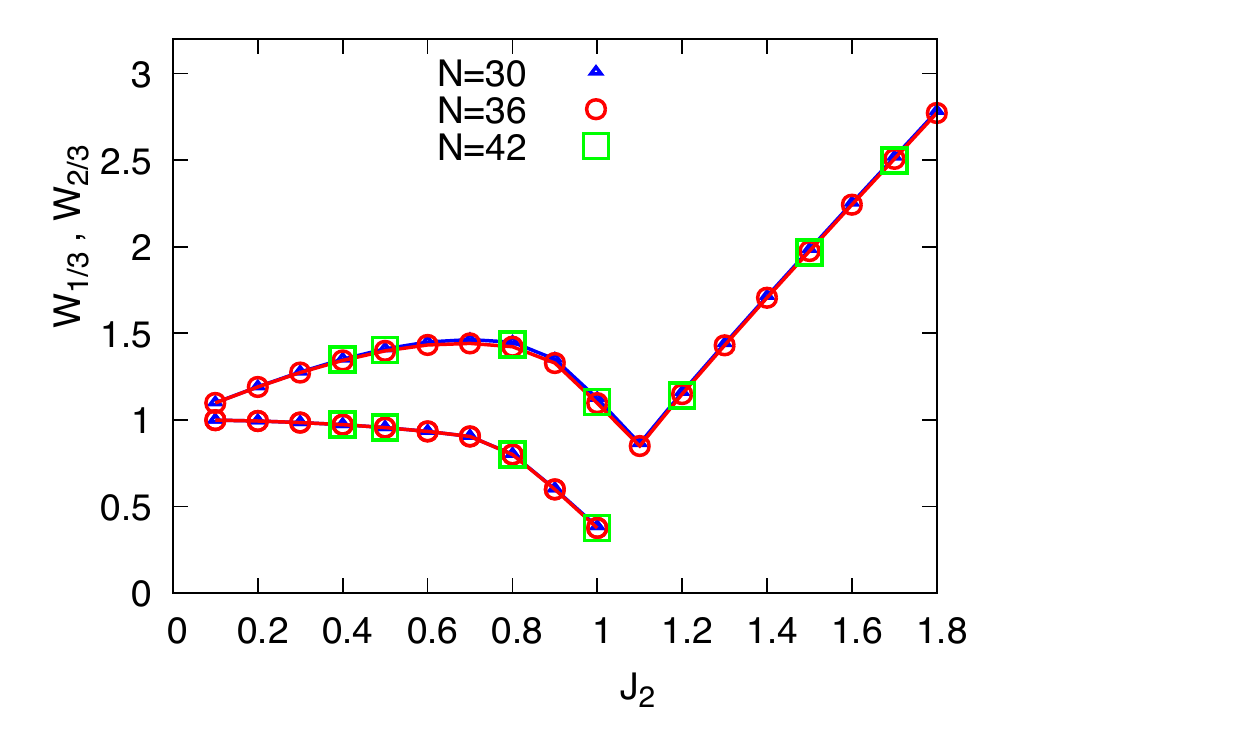}
\caption{Widths $W_{1/3}=g\mu_B(B_{2,1/3}-B_{1,1/3})$ of the $1/3$ plateau
and $W_{2/3}=g\mu_B(B_{2,2/3}-B_{1,2/3})$ of the $2/3$ plateau 
of the 
$J_1$-$J_2$ SKHAF  for $N=30$, $N=36$ and $N=42$. }
\label{fig_width}
\end{figure}

\begin{figure}[ht!]
\centering
\includegraphics*[clip,width=1.2\columnwidth]{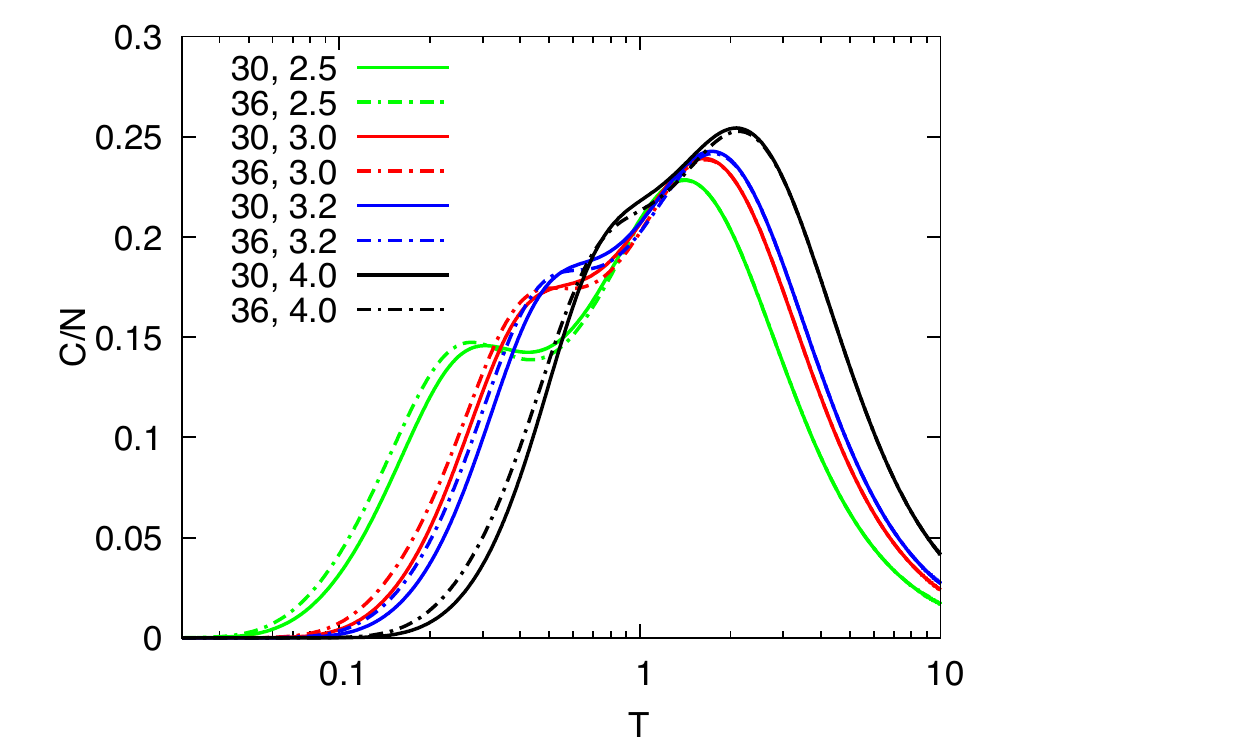}
\caption{Specific heat per site $C/N$ for $N=30$ (dashed lines) and $N=36$
(solid lines) for large values of $J_2$.
Note that in a wide temperature range the corresponding curves for $N=30$
and $36$ coincide.
}
\label{fig_C_large_J2_app}
\end{figure}

\begin{figure}[ht!]
\centering
\includegraphics*[clip,width=0.9\columnwidth]{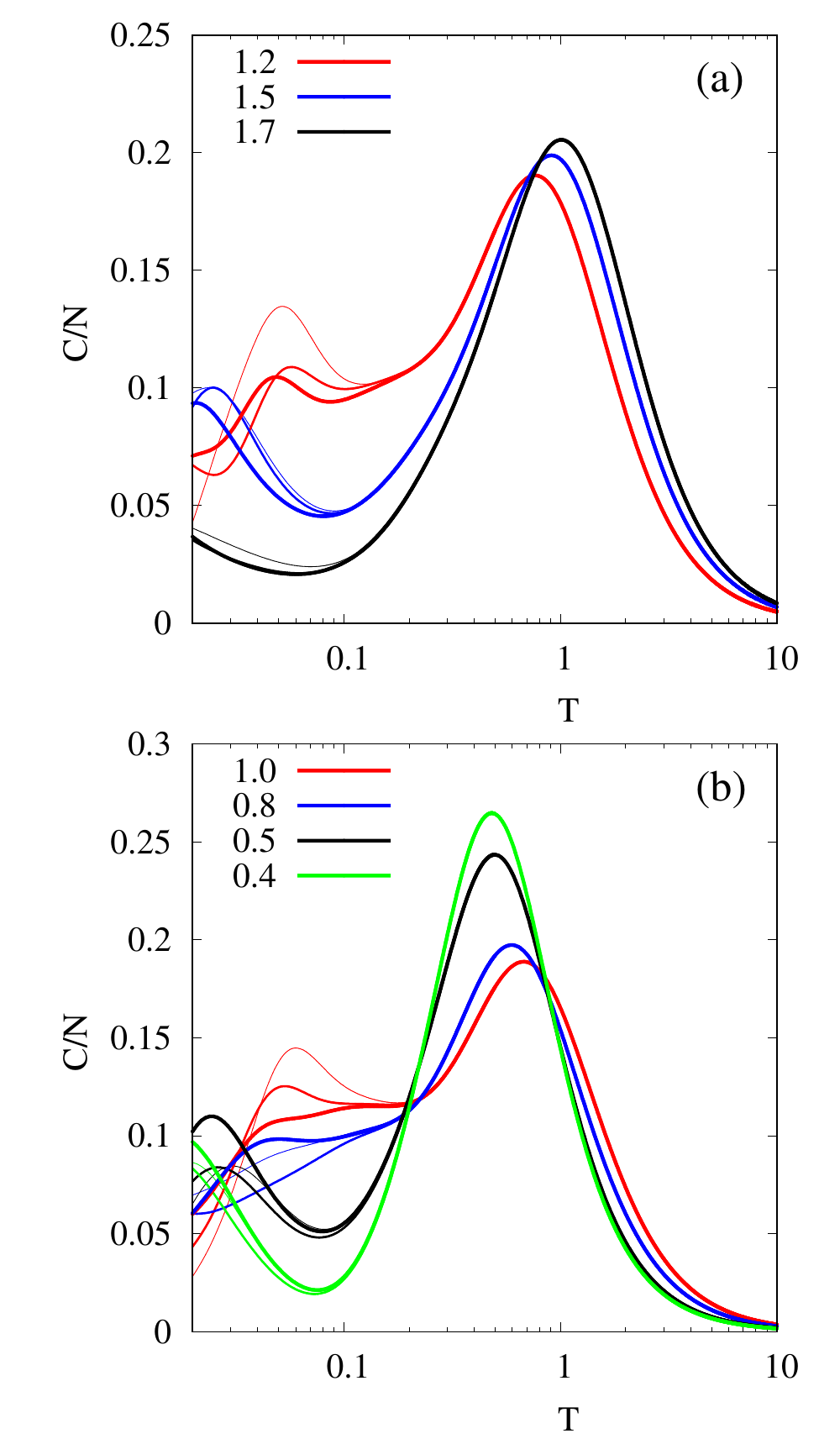}
\caption{Specific heat per site $C/N$ for $N=30$ (thin), $N=36$
(middle) and $N=42$ (thick). (a)  $J_2 > 1.0$,
(b) $J_2 \le 1.0$.
Note that in a wide temperature range the corresponding curves for $N=30$, $36$
and $N=42$ coincide.
}
\label{fig_C_app}
\end{figure}

\begin{figure}[ht!]
\centering
\includegraphics*[clip,width=0.9\columnwidth]{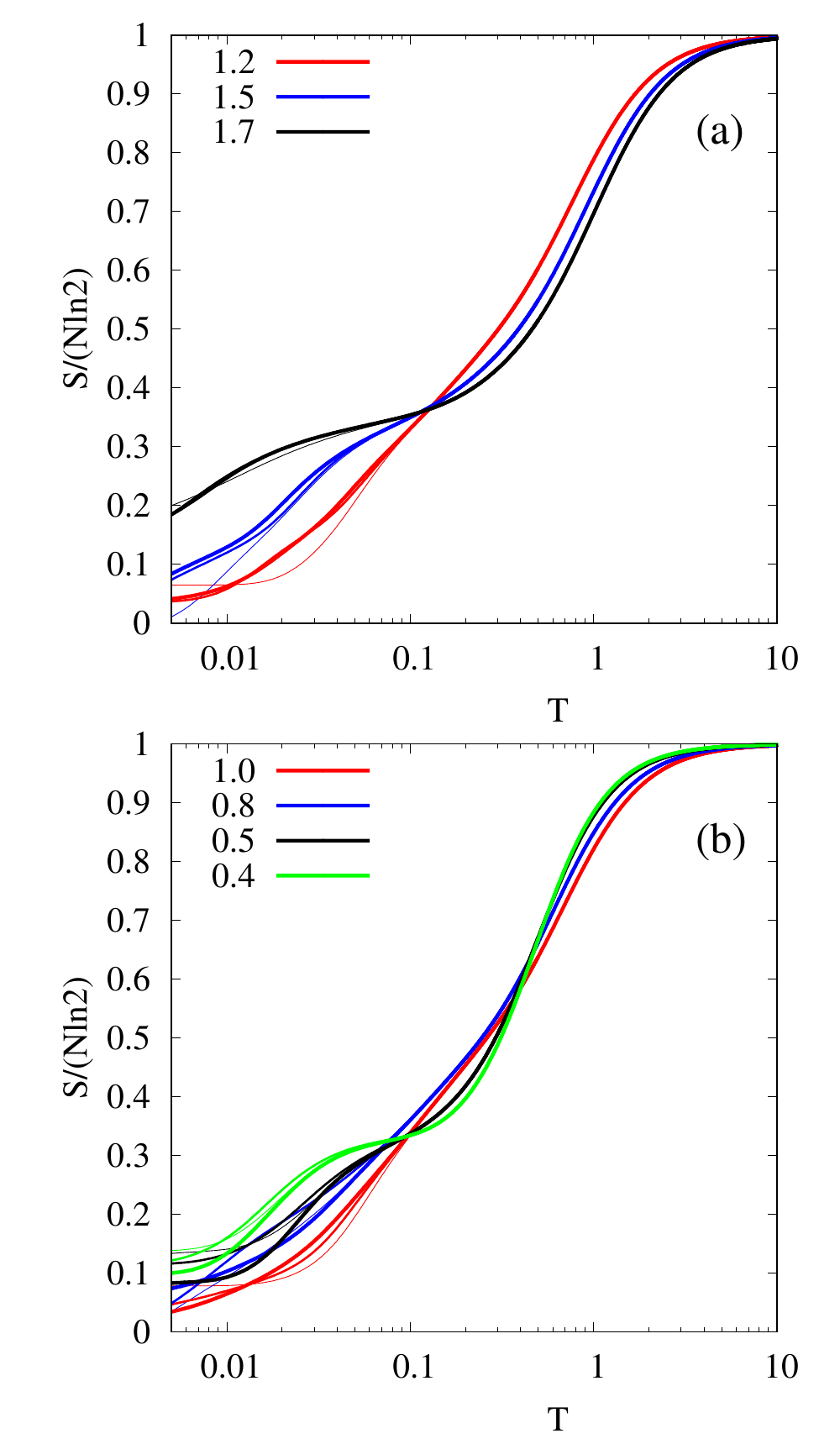}
\caption{Entopy per site $S/N$ for $N=30$ (thin), $N=36$
(middle) and $N=42$ (thick). (a)  $J_2 > 1.0$,
(b) $J_2 \le 1.0$.
Note that in a wide temperature range the corresponding curves for $N=30$, $36$
and $N=42$ coincide.
}
\label{fig_S_app}
\end{figure}

\begin{figure}[ht!]
\centering
\includegraphics*[clip,width=0.9\columnwidth]{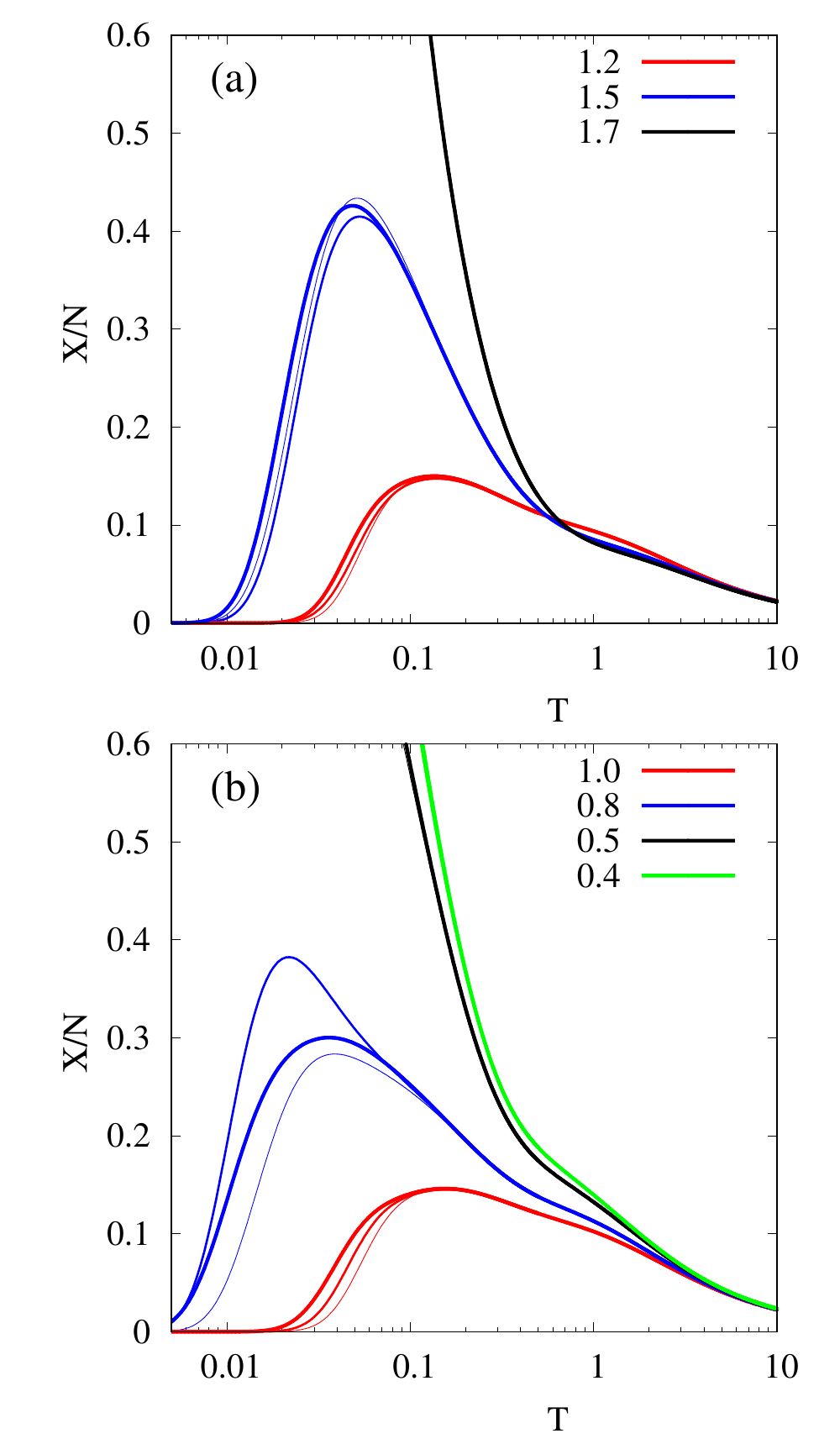}
\caption{Susceptibility per site $X/N$ for $N=30$ (thin), $N=36$
(middle) and $N=42$ (thick). (a)  $J_2 > 1.0$,
(b) $J_2 \le 1.0$.
Note that in a wide temperature range the corresponding curves for $N=30$, $36$
and $N=42$ coincide. Only for  $J_2 = 0.8$ there is a noticeable difference
around the maximum which is, however, at pretty low $T$.
}
\label{fig_chi_app}
\end{figure}

Here we present data  for the widths of the $1/3$ and $2/3$ plateaus
(Fig.~\ref{fig_width}) and show the specific heat for large values of $J_2$ 
(Fig.~\ref{fig_C_large_J2_app}).
Moreover, we provide additional information on finite-size effects for the specific heat
$C(T)$  (Fig.~\ref{fig_C_app}), the entropy $S(T)$ (Fig.~\ref{fig_S_app}) and the 
susceptibility $X(T)$ (Fig.~\ref{fig_chi_app}) by comparing data for $N=30$,
$N=36$, $N =42$.

\clearpage
\bibliography{JR_RGM,JR_own,js-own,js-other,allerlei,sawtooth}

\end{document}